\documentclass[aps,prl,twocolumn,reprint,showpacs,groupedaddress]{revtex4-1}
\usepackage{amsmath,amssymb,graphicx,color}

\usepackage{times}
\usepackage{latexsym}
\usepackage{stmaryrd}
\usepackage[table]{xcolor}
\usepackage{hyperref}
\usepackage{amsfonts}
\usepackage{bm}
\usepackage{bbm}
\usepackage{color}
\usepackage[english]{babel}
\usepackage{stmaryrd}
\usepackage{graphicx}
\usepackage{subfigure}
\usepackage{natbib}
\usepackage{appendix}
\usepackage{comment}

\newcommand\ket[1]{\left|\textstyle{#1}\right\rangle}

\newcommand{\mbp}[1]{{\color{black} #1}}
\newcommand{\mbptwo}[1]{{\color{black} #1}}

\newcommand{\jr}[1]{{\color{black} #1}}

\begin{document}
\bibliographystyle{unsrt}
\title{Blueprint for nanoscale NMR}
\author{I. Schwartz$^{1,3}$, J. Rosskopf$^{1}$, S. Schmitt$^{2}$, B. Tratzmiller$^{1}$, Q. Chen$^{1}$, L.P. McGuinness$^{2}$, F. Jelezko$^{2}$ and M.B. Plenio$^{1}$}
\affiliation{$^{1}$ Institute of Theoretical Physics and IQST,
Universit\"{a}t Ulm, 89081 Ulm, Germany \email{Corresponding authors: martin.plenio@uni-ulm.de \& ilai.schwartz@uni-ulm.de}\\
$^{2}$ Institute of Quantum Optics and IQST, Universit{\"a}t Ulm, 89081 Ulm, Germany\\
$^{3}$ NVision Imaging Technologies GmbH, 89134 Blaustein, Germany}

\begin{abstract}
Nitrogen vacancy (NV) centers in diamond have been used as ultrasensitive magnetometers to perform nuclear magnetic resonance (NMR) spectroscopy of statistically polarized samples at 1 - 100\,nm length scales. However, the spectral linewidth is typically limited to the kHz level, both by the NV sensor coherence time and by rapid molecular diffusion of the nuclei through the detection volume which in turn is critical for achieving long nuclear coherence times. 
Here we provide a blueprint for a set-up that combines a sensitivity sufficient for detecting NMR signals from nano- to micron-scale samples with a spectral resolution that is limited only by the nuclear spin coherence, i.e. comparable to conventional NMR.
Our protocol detects the nuclear polarization induced along the direction of an external magnetic field with near surface NV centers using lock-in detection techniques to enable phase coherent signal averaging. Using NV centers in a dual role of NMR detector and optical hyperpolarization source to increase signal to noise, and in combination with Bayesian interference models for signal processing, nano/microscale NMR spectroscopy can be performed on sub-millimolar sample concentrations, several orders of magnitude better than the current state of the art.
\end{abstract}
\maketitle

{\em Introduction ---}
Nuclear magnetic resonance (NMR) and magnetic resonance imaging (MRI) are technologies
whose applications in organic chemistry, biology, medicine and material science have
enabled fundamental scientific breakthroughs and continue to be drivers of scientific
and technological progress \cite{FindeisenB2014}. Despite these successes, it is recognized
that nuclear magnetic resonance applications has limitations due to the minute nuclear
magnetization of analytes which leads to limited sensitivity in comparison to other
analytic techniques such as mass spectrometry.

Strategies that are being pursued to overcome this challenge include an evolution towards
larger applied magnetic fields which improves sensitivity due to the resulting increase
of thermal equilibrium polarization and signal frequency \cite{BadilitaMS+2012}. The
approximately linear growth in the magnetic field that has been achieved over the last
$5$ decades comes at the cost of growing size, purchase and operating costs of these
devices, which limit portability and challenge their integration with desired applications.
More compact magnets lead to smaller usable detection volumes and thus limit sensitivity.
A promising alternative strategy is the reduction in size of the radio frequency
coils used to excite and detect the NMR signals \cite{Webb1997} as this results in a
sensitivity enhancement with decreasing coil-diameter and promises the development of
portable on-chip NMR spectrometers \cite{ZalesskiyDB+2014}. Limitations and challenges
in this approach include the homogeneity of the system which limit resolution and the
thermal noise in the readout coil, i.e. thermal Johnson noise, which, together with the
low sample volume, limits sensitivity. A further avenue towards improved NMR sensitivity
is to increase the nuclear spin polarization beyond its thermal equilibrium value by means of
\mbptwo{techniques such as} dynamical nuclear polarization \cite{ArdenkjaerLarsenFG+2003}. Despite promising results, the integration of
these approaches with NMR involve significant challenges, as they typically require low temperatures and dissolution of
the sample - significantly reducing its concentration.

Addressing these challenges in a single device to simultaneously achieve improved sensitivity,
ideally at the sub-millimolar level, portability and the ability to vary sample volumes
from the nano- to the millimeter scale would decisively enhance a broad range of applications
thus offering the potential for new ground breaking insights. These include NMR studies of
single cells and neurons \cite{GrisiVV+2017}, the study of catalysis at smallest volumes, NMR
studies of surfaces and on-chip NMR based metabolic fingerprinting with applications in
personalized medicine \cite{MarkleyBE+2017,Spiess2017}.


Here we present a novel physical platform for NMR detection that 
we show is capable of overcoming these challenges for samples ranging from the nano- to the millimeter-scale. We
introduce an NMR protocol that permits spectroscopy of such volumes with chemical resolution
and micromolar sensitivity and demonstrate signal processing algorithms that allow for a
significant reduction in signal acquisition time, thereby yielding sample analysis with
dramatic speed-up. The feasibility of this approach is demonstrated by detecting magnetic signals applied to a single NV center in diamond. The signal dynamics are obtained from atomistic simulations of a diffusive nanoscale nuclear sample, with a signal intensity scaled to correspond to an NV depth of 6.2 nm.


\begin{figure}[]
	\includegraphics[width=\columnwidth]{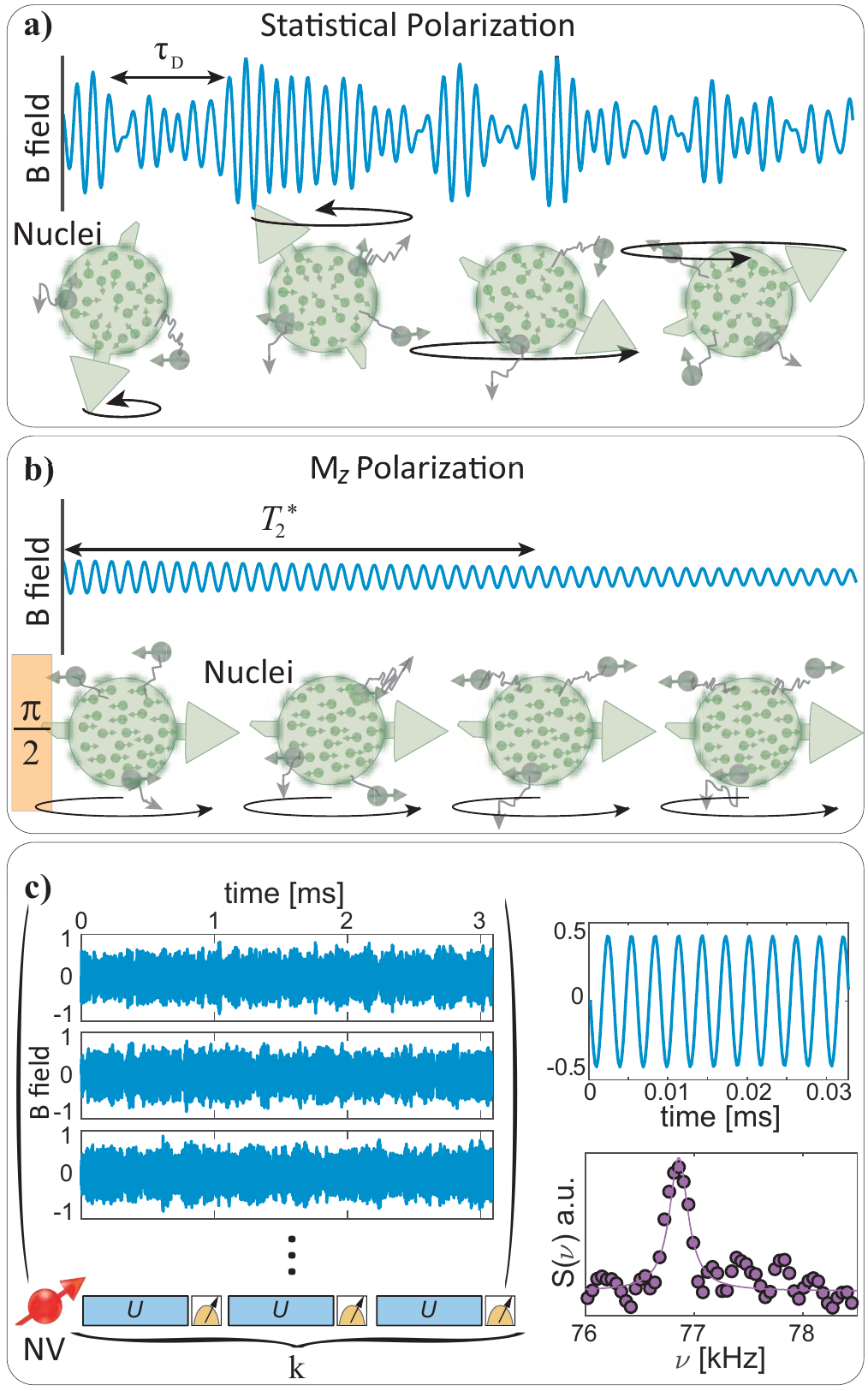}
	\caption{(a) Illustration of the Qdyne signal and the detected statistical magnetization by the NV center. The diffusion of nuclear spins in or out of the NV detection range leads to rapid fluctuations of the detected phase, introducing a short signal coherence time-scale, $\tau_D$, on the order of the interaction correlation time. (b) The adoption of $M_z$ Qdyne enables the detection of the FID signal from the nuclear z magnetization (thermal or hyperpolarized), regardless of the molecular diffusion. (c) An atomistic simulation of the time-dependent magnetic field induced by the diffusing nuclear spins on a 6.2 nm deep NV, where the statistical polarization is still clearly stronger than the hyperpolarized $M_z=0.1 \%$ polarization. However, $M_z$ Qdyne can be averaged over many runs $N_m$, significantly reducing the statistical polarization signal by $1/\sqrt{N_m}$ while the $M_z$ signal remains unchanged (due to same initial phase at every run). The figure on the top r.h.s. shows the averaged signal for the first 0.03 ms with $N_m=300$, where only the smaller $M_z$ polarization (oscillates between $\pm0.45$) remains visible. Using an AWG, the simulated magnetic field ($N_m = 1000$) was applied on a single NV center by a current carrying wire, with the setup of Ref~\cite{SchmittGS+2017}, where the $M_z$, diffusion independent signal can be clearly seen, producing a peak with 170 Hz linewidth (limited only by the 5.6 ms length of the simulated signal). A simulated signal with only statistical polarization produces no peak, and the diffusion limited signal would be over 10 kHz.}
	
	\label{sequences}
\end{figure}

{\em Background and key design elements ---} High-resolution NMR spectroscopy makes use of several properties of bulk matter in resolving chemical shifts and J-couplings for molecular structure determination. The rapid molecular diffusion and rotation leads
to the suppression of internuclear interaction down to the Hz-level while not limiting the
signal coherence due to the large volume from which the signal is collected. Furthermore,
for bulk samples, the thermal polarization (scaling with the sample volume V) greatly exceeds
the statistical polarization fluctuations (scaling with $V^{1/2}$). This allows on the
one hand for the controlled initialization of the signal and therefore phase coherent
signal accumulation resulting in a \mbptwo{rapid growth of} the signal to noise ratio (SNR) and on the other hand long signal coherence
times and therefore high spectral resolution.

However, for $(1\,\mu m)^3$ of water in a 1 Tesla field, the statistical polarization
of the hydrogen nuclei is comparable to the thermal polarization and becomes dominant at the
nanoscale. \mbptwo{This} observation has motivated successful experimental efforts towards
NMR detection of statistical polarization of nanoscale samples \cite{StaudacherSP+13}. \mbptwo{However,} 
the stochastic nature and random phase of the observed statistical polarization prevent phase coherent
signal averaging and the impact of diffusion \mbptwo{limits} signal coherence time and thus spectral
resolution~\cite{xi2015}. \mbptwo{Overcoming these limitations calls} for new modes of observation.

Here we will consider, perhaps counter-intuitively, the detection of the signal originating
from the thermal polarization, even for nanoscale samples for which the statistical component
is expected to dominate. We use three key features to compensate for this apparent shortcoming,
namely, (i) the signal phase can be controlled by an initializing $\pi/2$-pulse to allow for
phase coherent accumulation across subsequent measurements, (ii) because the thermal polarization
component is \mbptwo{uniform} across the entire sample beyond the immediate detection region,
the signal coherence time becomes essentially independent of diffusion allowing for high spectral
resolution, (iii) and for the same reasons the signal is uniform across the sample which allows
for the use of multiple NV-centers for simultaneous signal acquisition thus further improving SNR.


The platform and protocols described in the following leverage the unique characteristics of color
centers in diamond \cite{WuJP+2016} to make use of these three key features. First, we
use optically detected magnetic resonance in either individual or ensembles of color centers
\cite{GruberDT+1997,vanOortSG1990} to detect small magnetic fields emanating from the sample
which, by making use of a recently developed lock-in technique, Qdyne, which allows for
spectral resolution in the $\mu$Hz
range \cite{SchmittGS+2017,BossCZ+2017,BucherGL+17}. This substitutes the electrical detection via
rf-microcoils which is accompanied by thermal Johnson noise by optical detection which
is only limited by the non-thermal photon shot noise. Secondly, the ability to bring color
centers to within nanoscale distance of the sample allows for their use as a source of
nuclear hyerpolarization even under ambient conditions using laser induced polarization
of the electron spin native to the color center and the subsequent microwave assisted
transfer to the sample nuclei \cite{ChenSJ+2016,Fernandez-AcebalSS+2017,AbramsTE+2014}.
This obviates the need for a strong magnetic field and holds the potential for an orders
of magnitude increase in signal strength, thus bringing sub-millimolar sensitivities into
reach while at the same time reducing nuclear spin polarization fluctuations induced by
diffusion. Thirdly, in order to reduce averaging times required for achieving \mbptwo{sub-}millimolar
sensitivities, we employ signal processing methods based on Bayesian inference algorithms
that allow for orders of magnitude reduction of measurements required for the identification
of signal components due to chemical shifts.
%
The remainder of this work will describe these key elements and present \mbptwo{theoretical
and experimental results} that demonstrate the feasibility of the approach.

{\em Qdyne for nuclear magnetic resonance --- } The recently developed Qdyne method introduces
a quantum lock-in spectroscopy technique \cite{SchmittGS+2017,BossCZ+2017,BucherGL+17} whose
spectral resolution is independent of the sensor coherence time. Using Qdyne, a coherent external 
oscillating radiofrequency (RF) field, could be measured with a spectral linewidth of 607 $\mu$Hz 
\cite{SchmittGS+2017}, thereby making the technique promising for realizing true nanoscale NMR via 
shallow NV centers.

In Qdyne, the sensor qubit is tailored to collect a signal that depends not only on the amplitude
and frequency of the detected field, but also on the phase with respect to the start of each measurement.
Performing $N$ measurements each of length $T_L$, a different phase is accumulated in each measurement
due to the difference between $T_L$ and period of the oscillatory field. 
As shown in Ref.~\cite{SchmittGS+2017}, for an XY8 measurement sequence when the excitation and 
detection $\pi/2$ pulses are perpendicular to each other, the detected signal is given by
\begin{equation}
    P = \sin(\frac{4k\tau_m}{\pi}\cos(\delta t + \phi)) + \frac{1}{2},
\label{sig}
\end{equation}
where $k$ is the interaction strength, $\tau_m$ the interaction time, $\phi$ is an arbitrary initial
phase of the RF field, \mbptwo{and $\delta$ denotes} the frequency of the accumulated phase.

As with most NV sensing schemes, when using a \mbp{shallow NV for detection, due to the small number
$N_I$ of spins in the vicinity of the NV, the signal detected by Qdyne is dominated by the} statistical
polarization of the nuclear spins in the sensing volume - $B_{rms}(t) = \sum_i A^i_x(t) I_x$, with
$A^i_x(t)$ denoting the coupling of the NV center to the $i$-th nuclear spin.

Diffusion of molecules into and out of this volume leads to random fluctuations of the detected signal, which is governed by a correlation function $\langle B_{rms}(t)B_{rms}(t+\tau)\rangle = B_{rms}^2\exp(-\tau/\tau_c)$, where 
$\tau_c$ is the correlation time. Thus, the phase $\phi$ in Eq.\eqref{sig} becomes a stochastic 
variable, $\hat{\phi}(t)$, which denotes the instantaneous phase of the statistical nuclear spin polarization 
within the NV detection region \cite{Footnote1}.

This stochastic variable ties the observed Qdyne signal to the molecular diffusion of the moving
molecules, leading to a stringent limitation on the minimal observable line-width, thereby 
obscuring small but important details such as chemical shift and quadrupole information. See figure
1(a) for an illustration of a statistical magnetization detected by a Qdyne measurement, and the diffusion effect on the phase of the detected signal.

To solve this issue, and decouple the Qdyne signal from molecular diffusion, we modify the sequence to detect thermal nuclear magnetization along the z-axis of the applied magnetic field. This modified $M_z$ Qdyne sequence consists of $N_m$ measurements, and, at the beginning of the sequence, a $\pi/2$ pulse which rotates the nuclear z magnetization to the x-y plane,
where it can be detected by the XY dynamical decoupling measurement with the correct filter
function, similar to a free induction decay (FID) in traditional NMR. Importantly, at the beginning of
each of the $N_m$ sequences, the initial phase of the z magnetization on the x-y plane is
known and \emph{identical}. Moreover, as the phase is identical for all nuclear spins across the sample, the diffusion of molecules in or out of the NV center detection region has
no effect on the signal phase, and the detected linewidth becomes
limited only by the nuclear coherence times, see Fig.~\ref{sequences}(b).

In a realistic NMR scenario using a 5 -- 100\,nm deep NV center, both the statistical and $M_z$ sample magnetizations contribute to the detected signal, with the statistical polarization generally dominating the z-magnetization. However, the summation of $N_m$ repeated measurements can be used to reduce the statistical signal by a factor of $1/\sqrt{N_m}$ due to its random phase and magnitude, while locking-in to the phase of $M_z$-magnetization so as to prevent its cancellation. Thus, especially when combined with hyperpolarization, 
the detection of $M_z$ magnetization is feasible even with relatively shallow NV centers, and therefore even sub-micron detection volumes. Figure~\ref{sequences}(c)
\mbptwo{shows} the magnetization produced by an atomistic simulation of diffusing hyperpolarized ($<M_z> = 0.1\%$) nuclear spins \cite{Footnote2}
near a $6.2$ nm deep NV center. Clearly the statistical polarization is larger \mbptwo{than the 
hyperpolarized signal} in this regime as the FID is not visible. \mbptwo{However,} when 
averaging the signal over $300$ runs, the smaller $M_z$ polarization
can be clearly seen due to the reduction of the \mbp{contribution from the} statistical polarization.
The magnetic field produced by an ensemble of nuclear spins as calculated by atomistic simulations was applied to a single NV center by a current carrying wire (see Ref.~\cite{SchmittGS+2017} for experimental details) for $N_m = 1000$. The amplitude of
the signal is calibrated by fitting the accumulated phase due to the statistical polarization to that
measured by a $6.2$ nm deep NV center~\cite{Pham2016}. A measured signal due to the $M_z$ polarization
is clearly visible with $~170$ Hz linewidth, limited only by the 5.6 ms length of the detected signal.
For comparison, with the chosen diffusion parameter the linewidth due to the statistical polarization
would \mbptwo{exceed} $10$\,kHz.

An additional advantage of $M_z$ Qdyne due to the averaging of $N_m$ measurements is better
statistical information on each measured point. This improves the low photon collection efficiency
inherent in NV-based detection, with the statistical detection process becoming a Poissonian
distribution rather than a Bernoulli process. Moreover, as \mbp{the $M_z$-polarization is
uniform across the entire sample, different NV centers now detect the same phase $\phi$,
which} allows the measurement to be performed with ensembles of NV centers. The accumulated
fluorescence from the different NV centers acts identically to repeated measurements $N_m$ of
the single NV center, i.e. improving statistical information and averaging out the statistical
polarization. Thus, the number of \mbptwo{statistical averages is given by $N = N_{NV}\times N_m$,} 
where $N_{NV}$ is the number of NV centers used as sensors.

\begin{figure}
\includegraphics[width=\columnwidth]{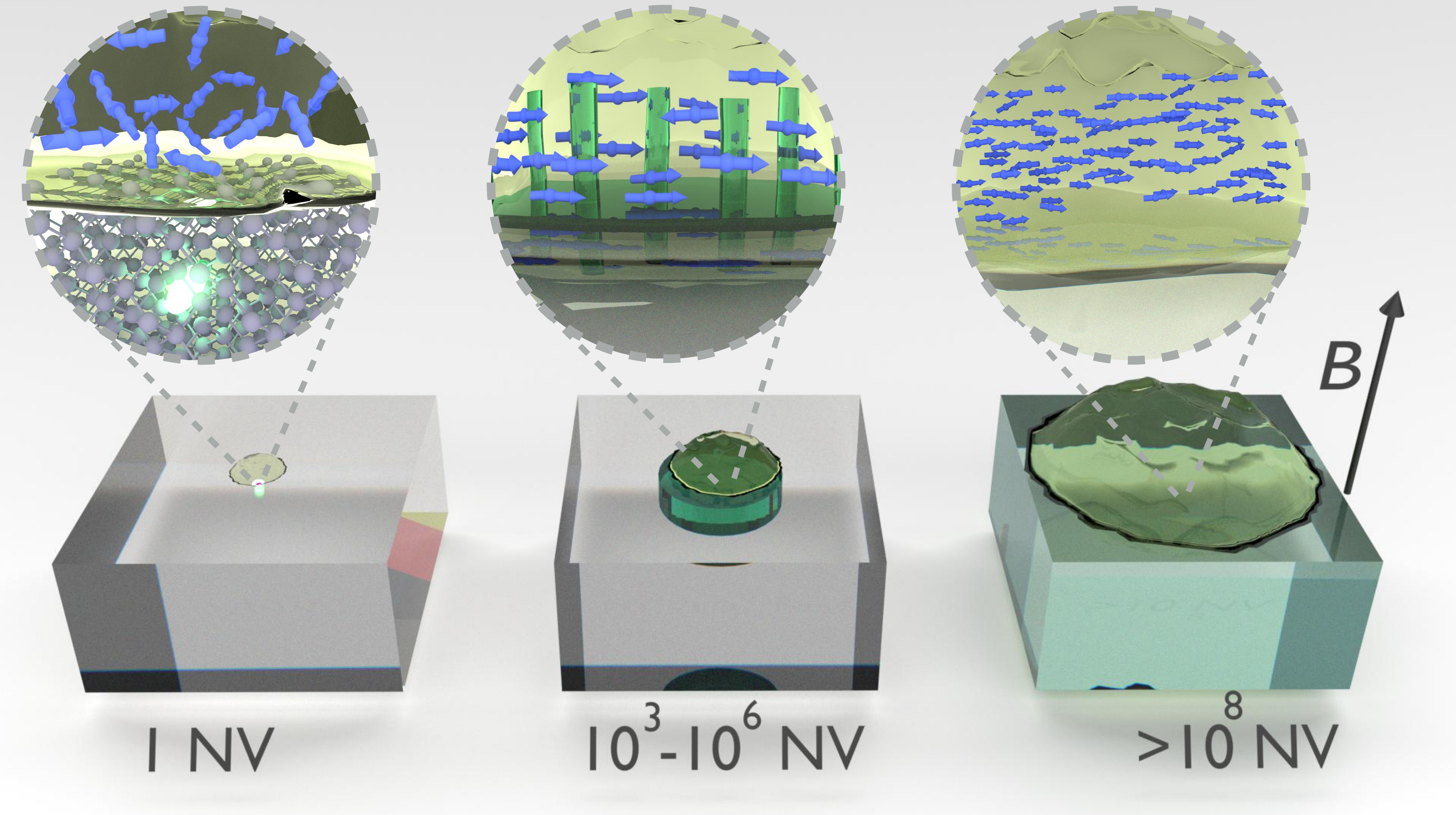}
\caption{Illustration of the three different regimes made possible by $M_z$ Qdyne.  On the r.h.s. is the ``classical'' regime, where a macroscopic diamond with densely packed NV centers
senses the thermal polarization, providing a substitute to traditional NMR micro-coils. In the middle figure, using the NV center ensemble in a dual role
- polarizing the nuclear spin bath to increase the nuclear signal and detection of the NMR signal,
thus termed Hyperdyne, enables superb sensitivities in the nano- and micro-scale regimes (the illustration includes a nanostructured diamond for enhanced polarization
efficiency). Hyperdyne can be pushed to the extreme nano-scale limit, using a single NV center (l.h.s.) albeit at a significantly reduced polarization / sensitivity efficiency.}	
	\label{regimes}
\end{figure}

$M_z$ Qdyne unlocks the potential of NV-based NMR, allowing for volumes ranging from nanometric
to macroscopic scales. However, as the nuclear thermal polarization is very weak, especially at
lower magnetic fields, achieving a good signal to noise (SNR) in this regime requires macroscopic diamonds with densely packed NV
centers and roughly 1 $\mu l$ samples (1 mm$^3$) for practical NMR applications (r.h.s. of fig 2). Pushing the
limits of this regime to even large micro-scale would require averaging of several hours of
measurement time for acquiring a sufficient SNR, due to the small signal produced by thermal
polarization. Thus, in this regime the NV ensemble in the diamond serves as a ``classical''
macroscopic NMR sensor, similar to currently used micro-coils, albeit with the advantage of
different noise processes which could lead to improved sensitivity. To push the application into the micro- and nano-scale regimes, the $M_z$ polarization needs to be enhanced, in a manner which still allows fast repetitive measurements. Fortuitously,
optically polarized NV centers have been \mbp{demonstrated to be} superb polarization sources for
nuclear spin hyperpolarization either inside the diamond~\cite{king2015room,alvarez2015local,scheuer2016optically,scheuer2017robust} or in external molecules~\cite{ChenSJ+2016,Fernandez-AcebalSS+2017,AbramsTE+2014}. Thus by using the NV centers
in a dual role of hyperpolarization sources and NMR detectors, with shallow NVs used for polarization 
and deeper ones for detection, the NMR SNR of each measurement can be increased over 10,000-fold, 
without the need for shuttling between polarization and detection \mbptwo{zones and without suffering}
other detrimental side-effects of dissolution DNP (e.g. cooling the sample to $T=1K$, reduction of 
analyte concentration upon dissolution). Thus, using interleaved hyperpolarization / detection sequences 
on the NV centers (the Hyperdyne protocol) one can achieve true NMR applicability on the (sub)microscale, 
see middle of fig. 2.

On the extreme nanometric scale (l.h.s. of fig 2), one \mbp{may implement Hyperdyne NMR with a single 
NV center.} However, the single NV center needs to be close to the surface for sufficient efficiency 
in the hyperpolarization cycle, which leads to the disadvantage that many of the polarized nuclei will 
diffuse outside the small NV detection region, and thus produce a much smaller net gain in the NMR 
sensitivity.

{\em \mbp{Signal Analysis of Hyperdyne: From FFT to Bayesian inference} --- } \mbp{In this
section we provide expressions for the signal to noise ratio for large numbers of detection
events, either due to high detection efficiency or large number of phase coherent averages,
and then present Bayesian inference methods to improve signal detection for when these
conditions are not met to allow for a significant order of magnitude reduction in measurement
time.} For shot noise limited detection,
\begin{equation}
	SNR \propto \sqrt{N_\text{Phot}} k \tau_m \propto \sqrt{N}   \rho P_n \tau_m,
\label{snr}
\end{equation}	
with $N_\text{Phot} \propto V_s$ being the number of detected photons, $\tau_m$ the
\mbp{the length of a single XY measurement}, $V_s$ the total \mbp{detection} volume,
$k$($\tau_m$) the interaction strength (time) for an individual NV center, $N=N_m N_{NV}$
the number of independent measurements (which is the product of $N_m$ runs with $N_{NV}$ NV centers),
$\rho$ the nuclear spin concentration and $P_n$ the average polarization. 
\mbptwo{There are two regimes to be considered when aiming  to maximize SNR in a given total 
experiment time here. Firstly, when $\tau_m < \min(T^{NV}_2,\pi/4k)$ 
it is most advantageous to increase $\tau_m$ while keeping $N_m$ constant, i.e. increase
the time over which the signal is accumulated coherently. In this case the total measurement 
time $T=N_m\tau_m$ scales as the first case of eq. (\ref{scaling}). If however, $\tau_m = T^{NV}_2$ 
we cannot increase $\tau_m$ any further without suffering an exponential in $\tau_m$ loss 
in signal (for $\tau_m > \pi/4k$ we lose the ability to identify the phase). Hence we are 
reduced to increase $N_m$, that is averaging over independent runs. In this case the  
total measurement time $T=N_m\tau_m$ scales as the second case of eq. (\ref{scaling})
benefiting more from an increase in polarization of the sample. }

\begin{equation}
\frac{1}{T} \propto \begin{cases} \mbptwo{\sqrt{V_s\rho_{NV}}} \rho P_n&\mbox{if increasing } \tau_m < \text{min} \left(T^{NV}_2, \frac{\pi}{4k}\right)  \\
V_s\rho_{NV} \rho^2 P_n^2 & \mbox{else}. \end{cases}
\label{scaling}
\end{equation}
where $T$ is the total measurement time for achieving a fixed SNR value, $\rho_{NV}$ is the 
NV concentration, and enlarging $V_s$ is assumed to be \mbptwo{achieved by increasing} the 
surface cross section, thereby scaling linearly with the number of NV centers.

\begin{figure}
\includegraphics[width=\columnwidth]{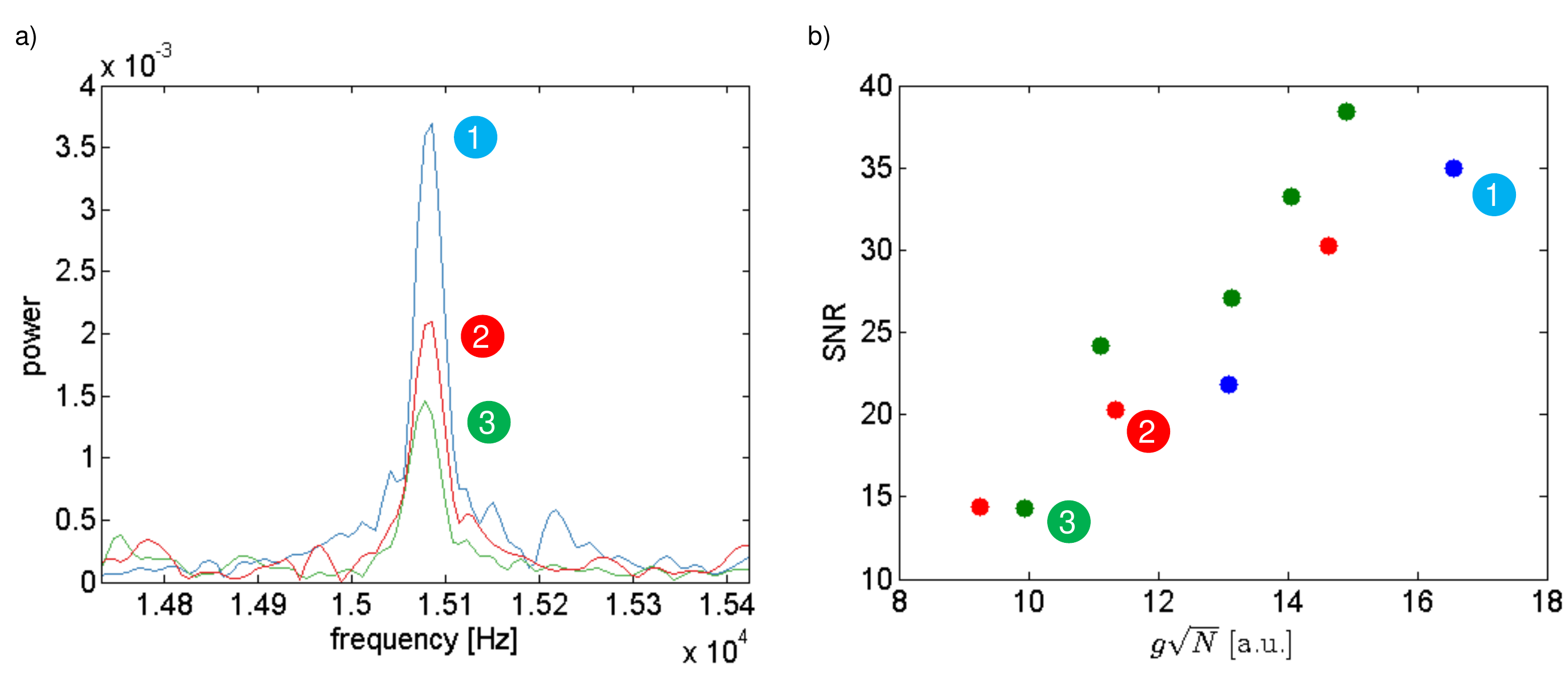}
\caption{(a) The Fourier transform of the NV Hyperdyne signal with three different parameters
for the diffusing nuclear spins.  The SNR mainly depends on the signal amplitude within each
XY measurement $g=k\tau_m$ and the number of measurements $N$. (b) Scaling of the SNR with
$g\sqrt{N}$ for numerous parameter configurations, noting specifically the three parameters
from (a), curve expected to  be linear by Eq.~\ref{snr}. \mbp{The parameters are $P_n=0.1\%,
\rho=5M, N = 1600, \tau_m = 4.1 \mu s$ (data set 1), $P_n = 0.5\%, \rho = 5M, N=30, \tau_m =
4.1 \mu s$, (data set 2) and $P_n = 0.05\%, \rho=150 mM, N=400, \tau_m = 32.8 \mu s$ (data
set 3).} } \label{scaling}
\end{figure}

Figure 3(a) shows the Fourier transform of the acquired $M_z$ Qdyne signal by a
\mbp{6.2} nm deep NV center for three scenarios with different polarization, molecular
concentration and number of measurements $N$. All scenarios were produced by atomistic
simulation of the detection process for \mbp{diffusing nuclei at the density of
water}. One can see a difference in the SNR between the three scenarios, due to
a difference in $g=k\tau_m$ and $N$. As shown in figure 3(b), the scaling of the
SNR is proportional to $g\sqrt{N}$, as expected from Eq.~\ref{snr}
\cite{Footnote3}.

As noted above, the statistical polarization can be larger than the $M_z$ polarization
as it is reduced by the averaging of the signal. The statistical polarization does 
limit the accumulation time in the XY sequences for shallow NV centers, as the 
condition $\gamma_e B_{rms} t_m < \pi/2$ needs to be fulfilled \mbp{to ensure that 
the $M_z$-signal is not fully randomized by the statistical polarization}. As hyperpolarization 
enhances the $M_z$ to statistical polarization ratio, a larger $M_z$ signal can 
be accumulated by shallow NV centers, enabling the sensing of nano-scale volumes. 
\mbp{It is important to note that the shot noise in the detection process scales as $1/\sqrt{N}$, 
as does the statistical polarization signal, which in turn implies that the 
fluctuations due to the statistical polarization signal are never larger than the 
shot noise, and are typically negligible}.

It is interesting to note the comparison to microcoils. When scaling the diamond to the macroscopic 
regime (e.g. 1 $\mu$l), the expected sensitivity will be similar to that achieved with state of the 
art micro-coils. However, due to the ohmic contribution to the noise becoming dominant at small 
diameters~\cite{Peck1995}, microcoil sensitivity per unit volume starts scaling as $1/\sqrt{d}$ 
instead of $1/d$~ when $d<100 \mu$m \cite{Webb1997}, where $d$ is the coil diameter, equivalent 
to $1/V_s^{1/4}$. Therefore, even for large micrometric samples NV detection starts 
\mbp{may become superior to} microcoil detection, even without combining with hyperpolarization.

The measured Hyperdyne photon count signal $\bf{D}$ is acquired by photo detectors \mbp{with a
detection scheme similar to that of \cite{gaebel2006} and suffers from several loss sources that
make the acquired signal very noisy.} On the one hand the detection is extremely lossy, leading
to a very sparse time series of photon counts \mbp{with less than a detection event per signal
period}. On the other hand each NV emits with a finite probability a photon in the $\ket{-1}$
state, leading to just a small net difference in detection probability $p$ between $\ket{0}$-state
($p_{\ket{0}} \approx 4.0\%$) and $\ket{-1}$-state ($p_{\ket{-1}} \approx 2.5\%$)~\cite{Footnote4}. 
In the framework of Fourier NMR spectroscopy, this setting leads to rapidly decaying SNR.

In this challenging, low SNR settings, approximating the parameters of an underlying, hidden model 
by Bayesian inference has shown great benefit in other experimental scenarios (e.g. in astro- or 
particle-physics \cite{hilbe2017, feroz2009} and recently in NV center measurements \cite{Hincks2017}). Similar to FFT the Bayesian method operates on the 
raw signal vector $\bf{D}$ without any preprocessing or reconstruction, but at the same time reducing 
the measurement time by at least one \mbp{order of} magnitude in typical settings. It relies on a 
probabilistic graphical  model (PGM) capturing the hierarchical nature of the Rabi oscillation, photon 
emission and detection. This parametric model allows to incorporate prior knowledge of the problem into 
the analysis of the sparse signal. \jr{By using Bayesian inference NMR spectroscopy can be interpreted 
as fitting the distribution of parameters of an underlying harmonic model. The fit is guided by measured 
data $\bf{D}$ and an informed choice of priors of the parameters $\Theta$. The priors which go into the population probability $P$ in eq. \eqref{sig} are determined by a normally distributed 
$g \sim  \mathcal{N}_\mu(\mu=4k\tau_m/\pi)$, the uniform oscillation frequency 
$\delta \sim \mathcal{U}_{a,b}(\delta_0-a, \delta_0+b)$ and an uniform free phase parameter 
$\phi \sim \mathcal{U}_{a,b}([0, 2\pi])$. Descending from $P$ the measurement is modeled by 
$M \sim \mathcal{P}_\lambda (\lambda = p_{\textrm{dark}} + (p_{\textrm{bright}} - p_{\textrm{dark}}) P)$, 
where $\mathcal{P}_\lambda$ is the Poisson distribution for the photon counts. It's rate parameter 
$\lambda$ is determined by the parent emission process. The resulting distributions 
after the fit are called posteriors.}

$\bf{M}$ is a vector of stochastic random variables (RVs) as it depends on parents in the PGM. 
The parents could either be constants or random according to a specified probability distribution. The 
value of the vector is determined by the measured photon counts. It's up to the inference mechanism to 
estimate the posterior model parameters $\Theta = {g, \delta, \phi}$ such, that the posterior distribution 
approximates the measured values best. The adjustment of the posteriors is done by drawing many samples 
from a proposal distribution using Markov Chain Monte Carlo (MCMC) and either keeping the current $\Theta$ 
with a certain probability if the likelihood of the measured data is increased, or rejecting the sample. 
The MCMC sampling takes the form of a Markov-Chain which means the position of step $n+1$ is dependent 
only upon the position of step $n$, and is otherwise independent of all other steps. The walk around the 
joint proposal distribution happens in a semi-random manner. The step-size and direction are decided according 
to specific rules of the sampling method, including randomness (the Monte-Carlo aspect) and gradient-seeking 
and momentum (Hamilton Monte Carlo \cite{hoffman2014}) for efficiency. If the MCMC algorithm has converged 
sufficiently well, the samples drawn approximate the respective posterior distributions of the RVs. The 
whole procedure can be seen as a stochastic simulation of the experiment and adjustment of the parameters 
until measured and simulated data is \mbp{statistically} equivalent. To implement the inference algorithm we
relied on recent software techniques \cite{Footnote_PYMC}.

\mbptwo{The benefit of a Bayesian analysis can be seen in Fig 4b where, in this specific setting, 
the inset demonstrates that the Bayesian analysis allows for an order of magnitude reduction in the minimal detectable concentration.}

{\em Achieving sub-millimolar detection limit in the (sub)micro scale --- } Hyperdyne achieves
excellent SNR with a relatively small $N_m$. Thus, the \mbp{key} question is whether a diamond-based
setup can combine high nuclear polarization with $M_z$ Qdyne to achieve applicable NMR spectrometry
on the nano-micro scale.

\begin{figure}
	\includegraphics[width=\columnwidth]{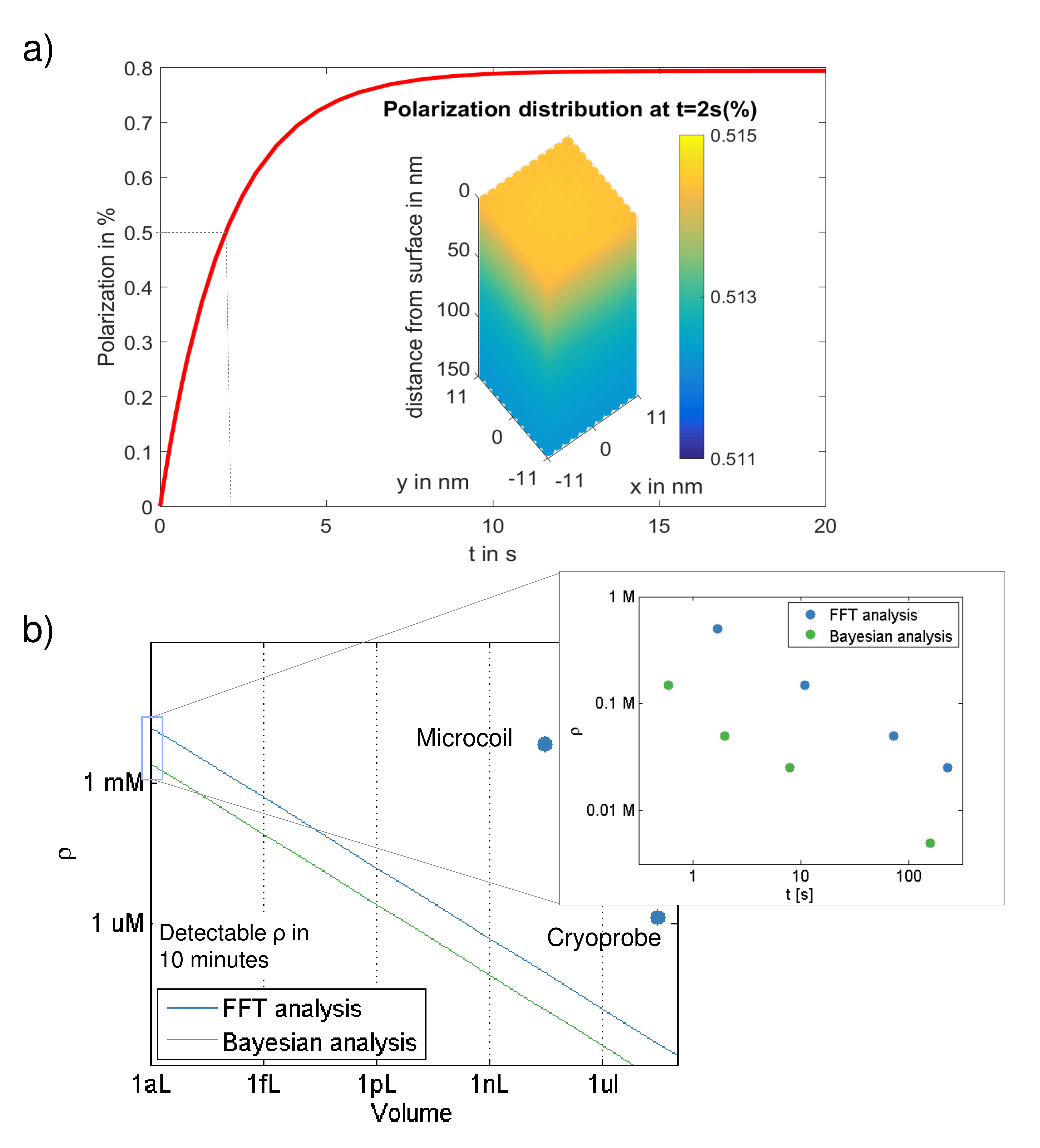}
	\caption{(a) Simulation of the polarization buildup for our setup, taking into account the polarization rate, molecular diffusion and nuclear relaxation of $T_1 = 2$ s. (b) The minimal detectable molecular concentration for different volumes using the $\sqrt{N}$ dependence (volume increased by increasing the surface cross-section). An SNR of 10 was chosen as the threshold. Also shown are the best achieved sensitivities for cryoprobes and microcoils~\cite{Kehayias2017}. The insert depicts the time required for detection (SNR $>$ 10) of different concentrations for a volume of $(100 nm)^3$, used for the calculation of the detectable minimal concentration within 10 minute time.}
	
	\label{regimes}
\end{figure}

We propose the following setup for nano-micro scale NMR (fig. 2(b))- an analyte with $\mu$M-mM 
concentration in a solution is placed on top of a nanostructured diamond for improved surface ratio, e.g.
nanoslits~\cite{Kehayias2017}. The $\sim0.3\mu m$ wide nanoslits are filled with NV centers with 
concentration of $10^{17} \, cm^{-2}$, including near the surface (5-20 nm deep). The Hyperdyne sequence
is then composed of $N_m$ alternations of hyperpolarization and $M_z$ Qdyne sequence. The 
hyperpolarization, driven by the shallow NV centers (5-20 nm deep) significantly enhances the measured signal.

Since NV centers can be optically polarized to over 90\% polarization with microsecond-long laser
pulses~\cite{Manson2006}, shallow NVs provide a unique resource for polarizing nuclear spins in nearby
molecules. The polarization efficiency depends on $g_{tot}\tau_c$, where $\tau_c$ is the correlation time, 
and $g_{tot}=g\sqrt{N_I}$ is the total flip-flop coupling between the nuclear spins and the NV center, $N_I$
the number of nuclear spins in the detection region, and $g$ the average coupling to these nuclear spins. 
When $g_{tot}\tau_c < 1$, the typical scenario, the polarization efficiency does not depend on the analyte
concentration~\cite{Fernandez-AcebalSS+2017}. 

As the distance between slits is 150 nm, and with the chosen NV concentration the average distance between NV
centers is ~22 nm, to calculate the achieved polarization, on needs to consider the polarization buildup in a
$22nm\times 22nm\times 150nm$ region for each NV. Assuming $g_{tot}\tau_c < 1$, nuclear $T_1 \sim 2$ s, using
robust polarization pulses \cite{SchwartzST+2017} and experimentally verified polarization transfer
rates~\cite{Fernandez-AcebalSS+2017}, we obtain 0.5\% polarization of the analytes in the nanoslit solution
after 2 seconds of polarization, as shown in figure 4(a). the diffusion of the analyte is assumed to be around
$10^{-11} m^2/s$, either due to large analytes (e.g. proteins) or a viscous solution. The polarization
simulation takes into account the polarization rate, the nuclear relaxation process and the molecular
diffusion.

For thermal polarization the duty cycle for the measurements is limited by the time required to build up the
thermal polarization, on the order of several $T_1$ times. Unsurprisingly, as the nuclear $T_1$ is the limiting
time also for the polarization buildup, replacing nuclear thermalization by NV-based DNP does not change the
duty cycle of the NMR experiments.

The achieved NMR sensitivity now depends on the volume of material probed, as $N_{NV}$ scales linearly with the
volume (the solution is assumed to reside mainly in the nanoslits, but the diamond surface cross section can be
enlarged). Fig. 4(b) shows the detectable analyte concentration within 10 minutes of measurement time (taking
into account the time for polarization and the experimental photon detection efficiency) when varying the probe
size, for the achieved polarization of $0.5\%$ in each hyperpolarization cycle, with spectral and Bayesian
analysis. As a comparison, best achieved sensitivity for microcoil and cryogenic probe NMR are
noted~\cite{Kehayias2017}. For NMR spectroscopy we see that for a volume of less than 1 femtoliter (1
nanoliter), detection with a few Hz resolution of mM ($\mu$M) concentrations is feasible with Hyperdyne within
10 minutes, corresponding to $3\times 10^6$ spins/$\sqrt{\text{Hz}}$ ($10^{18}$  spins/$\sqrt{\text{Hz}}$), paving the way for applicable diamond-based NMR spectroscopy. Even with hyperpolarization, these
regimes would not be possible with standard microcoils due to the $V^{1/4}$ scaling. Note that at very low
concentrations, additional noise will be produced by the spatial location of the individual number of nuclear
spins at the NV detection region.

Regarding the minimal linewidth detectable by the setup, due to the small width of the slits (300 nm), all
nuclear spins are within relatively close proximity of to the NV center sensors. Integrating the total
deviation to the nuclear Larmor frequency due to the Z-Z coupling with the NV centers leads to $\gamma_e\Delta
B \approx 1-2$ Hz. The presence of NV centers on both sides of the slit actually improves the homogeneity, as
the deviation of the Larmor becomes more homogeneious spatially\cite{supp}. This deviation might cause some
broadening in high-resolution NMR, but does not inhibit the acquisition of precise spectra to within $\sim$1 Hz
resolution.

{\em Discussion --} The ability to use NV centers as a hybrid quantum-classical detector together
with hyperpolarization \mbp{and signal processing based on Bayesian inference} opens up new possibilities
for (sub)micro scale NMR. It is interesting to note the uniqueness of the system - the NMR detection
is based on individual electron spins, accumulating phase independently, very differently from other
methods of NMR detection (e.g. induction in tuned coils). The fact that these same electron spins (at
least the shallow ones) can also be optically polarized and serve as a source for suprathermal
dynamic nuclear polarization for the investigated nuclear spins is a fortuitous coincidence for the
hybrid polarizer/micro-NMR system, and enables achieving remarkable sensitivities. The achieved
polarization of the molecules will depend on the molecular relaxation time and diffusion, and can
be optimized for specific molecules.

Regarding the analysis of the $M_z$ Qdyne signal, the introduced Bayesian analysis was shown to dramatically
improve the detection sensitivity. It is worth exploring how well this analysis could also improve non-
hyperpolarized Qdyne, as it could push the limits (concentration, time, volume) of the regimes where it is
applicable.

{\em Summary and Conclusions ---} In this work we have presented a blueprint for nanoscale NMR. Our approach
builds on earlier work that demonstrates experimental feasibility of the required magnetic field detection
scheme \cite{SchmittGS+2017}, on theoretical and experimental work that developed and demonstrated polarization
transfer from color centers to liquids \cite{ChenSJ+2016,Fernandez-AcebalSS+2017,AbramsTE+2014} and signal
processing methods \cite{flegal2011,hilbe2017}.

{\em Acknowledgements --- } The authors would like to thank Jochen Scheuer and Julen Simon Pedernales for
discussions and support. This work was supported by the EU projects HYPERDIAMOND and DIADEMS,
the ERC Synergy grant BioQ and a PhD fellowship of the Integrated Center for Quantum Science and
Technology (IQST).


\begin{thebibliography}{}
\bibitem{FindeisenB2014} M. Findeisen, S. Berger, {\em 50 and More Essential NMR Experiments:
A Detailed Guide}, Wiley-VCH,Weinheim, 2014.
%
\bibitem{BadilitaMS+2012} V. Badilita, R. Ch. Meier, N. Spengler, U. Wallrabe, M. Utz, and J.G. Korvink,
{\em Microscale nuclear magnetic resonance: a tool for soft matter research.} Soft Matt. {\bf 8}, 10583 (2012).
	%
\bibitem{Webb1997} A.G. Webb, {\em Radiofrequency microcoils in magnetic resonance.} Prog. Nuc. Magn.
Res. Spectrosc.	{\bf 31}, 1 (1997).

\bibitem{Spiess2017} H.W. Spiess, {\em 50th Anniversary Perspective: The Importance of NMR Spectroscopy to Macromolecular Science.} Macromolecules {\bf 50}, 1761 (2017).

	%
\bibitem{ZalesskiyDB+2014} S.S. Zalesskiy, E. Danieli, B. Blümich, and V.P. Ananikov, {\em Miniaturization
of NMR Systems: Desktop Spectrometers, Microcoil Spectroscopy, and “NMR on a Chip” for Chemistry, Biochemistry, and Industry.} Chem. Rev. {\bf 114}, 5641 (2014).
	%
\bibitem{ArdenkjaerLarsenFG+2003} J.H. Ardenkjaer-Larsen, B. Fridlund, A. Gram, G. Hansson, L. Hansson,
M.H. Lerche, R. Servin, M. Thaning, and K. Golman, {\em Increase in signal-to-noise ratio of > 10,000
times in liquid-state NMR.} Proc. Natl. Acad. Sci. USA. {\bf 100}, 10158 (2003).
	
\bibitem{DuckettM2012} S.B. Duckett and R.E. Mewis, {\em Application of Parahydrogen Induced Polarization
		Techniques in NMR Spectroscopy and Imaging.} Acc. Chem. Res. {\bf 45}, 1247 (2012).
	%
\bibitem{GrisiVV+2017} M. Grisi, F. Vincent, B. Volpe, R. Guidetti, N. Harris, A. Beck and G. Boero,
	{\em NMR spectroscopy of single sub-nL ova with inductive ultra-compact single-chip probes.} Sci. Rep.
	{\bf 7}, 44670 (2017).
	%
\bibitem{MarkleyBE+2017} J.L. Markley, R. Br{\"u}schweiler, A.S. Edison, H.R. Eghbalnia, R. Powers,
	D. Raftery and D.S. Wishart, {\em The future of NMR-based metabolomics.} Curr. Opin. Biotech. {\em 43},
	34 (2017).
	%
	%
\bibitem{StaudacherSP+13} T. Staudacher, F. Shi, S. Pezzagna, J. Meijer, J. Du, C.A. Meriles, F. Reinhard
and J. Wrachtrup, {\em Nuclear magnetic resonance spectroscopy on a (5-nanometer)$^3$ sample volume},
Science {\bf 339}, 561 (2013).

\bibitem{WuJP+2016} Y. Wu, F. Jelezko, M.B. Plenio and T. Weil, {\em Diamond Quantum Devices in Biology.}
	Angewandte Chemie – International Edition Minireview {\bf 55}, 6586 (2016).
	%
\bibitem{GruberDT+1997} A. Gruber, A. Dr{\"a}benstedt, C. Tietz, L. Fleury, J. Wrachtrup and C. von
	Borczyskowski, {\em Scanning Confocal Optical Microscopy and Magentic Resonance on Single Defect
		Centers in Diamond.} Science {\bf 276}, 2012 (1997).
	%
\bibitem{vanOortSG1990} E. van Oort, P. Stromer and M. Glasbeek, {\em Low-field optically detected
		magnetic resonance of a coupled triplet-doublet defect pair in diamond.} Phys. Rev. B {\bf 42},
	8605 (1990).
	%
\bibitem{SchmittGS+2017} S. Schmitt, T. Gefen, F.M. Stürmer, T. Unden, G. Wolff, Ch. Müller, J. Scheuer,
	B. Naydenov, M. Markham, S. Pezzagna, J. Meijer, I. Schwarz, M. B. Plenio, A. Retzker, L.P. McGuinness,
	and F. Jelezko, {\em Sub-millihertz magnetic spectroscopy performed with a nanoscale quantum sensor.}
	Science {\bf 351}, 832 (2017).
	%
\bibitem{BossCZ+2017} J.M. Boss, K.S. Cujia, J. Zopes, and C.L. Degen, {\em Quantum Sensing with arbitrary
		frequency resolution.} Science {\bf 351}, 837 (2017).
	%
\bibitem{BucherGL+17} D.B. Bucher, D.R. Glenn, J. Lee, M.D. Lukin, H. Park, and R.L. Walsworth, {\em
High Resolution Magnetic Resonance Spectroscopy Using Solid-State Spins.} E-print arXiv:1705.08887.
%
\bibitem{ChenSJ+2016} Q. Chen, I. Schwarz, F. Jelezko, A. Retzker and M.B. Plenio, {\em Resonance-inclined
		optical nuclear polarization of liquids in diamond structures.} Phys. Rev. B 93, 060408(R)(2016).
	%
\bibitem{xi2015} X. Kong, A. Stark, J.F. Du, L.P. McGuinness and F. Jelezko,
{\em Towards Chemical Structure Resolution with Nanoscale Nuclear Magnetic Resonance Spectroscopy.} Phys. Rev. Appl. {\bf 4}, 024004 (2015)
	
\bibitem{Fernandez-AcebalSS+2017} P. Fernandez-Acebal, O. Rosolio, J. Scheuer, I. Schwarz, B. Tratzmiller, Q. Chen,
	C. M{\"u}ller, B. Naydenov, A. Retzker, M.B. Plenio, and F. Jelezko. {\em Polarisation of oil molecules via nitrogen-vacancy
		centers in diamond.} In preparation.
%
\bibitem{AbramsTE+2014} D. Abrams, M.E. Trusheim, D.R. Englund, M.D. Shattuck, and C.A. Meriles,
	{\em Dynamic nuclear spin polarization of liquids and gases in contact with nanostructured diamond.}
	Nano Lett. {\bf 5}, 2471 (2014).
	%
\bibitem{Footnote1} Some fluctuations in \mbp{$\phi$} are still expected due to spatial deviations
of the molecule locations, however, this is expected to be negligible even for shallow NVs. $k$ 
would also fluctuate due to the density and spin fluctuations, but to a much lesser extent than 
$\hat{\phi}(t)$.
%
\bibitem{Footnote2} The diffusion coefficient was chosen to be $D=10^{-12} m^2/s$, similar to
oil molecules.
%
\bibitem{Pham2016}
	Linh M. Pham, Stephen J. DeVience, Francesco Casola, Igor Lovchinsky, Alexander O. Sushkov, Eric Bersin, Junghyun Lee, Elana Urbach, Paola Cappellaro, Hongkun Park, Amir Yacoby, Mikhail Lukin, and Ronald L. Walsworth
	\textit{NMR technique for determining the depth of shallow nitrogen-vacancy centers in diamond.}
	Physical Review B 93.4 (2016): 045425.
%
\bibitem{scheuer2016optically}
J. Scheuer, I. Schwarz, Q. Chen, D. Schulze-S{\"u}nninghausen, P. Carl, P. H{\"o}fer,
A. Retzker, H. Sumiya, J. Isoya, B. Luy, M.B. Plenio, B. Naydenov, and F. Jelezko.
{\em Optically induced dynamic nuclear spin polarisation in diamond.} New J. Phys.
{\bf 18}, 013040 (2016).
%
\bibitem{alvarez2015local}
G.A. {\'A}lvarez, C.O. Bretschneider, R. Fischer, P. London, H. Kanda, S. Onoda,
J. Isoya, D. Gershoni, and L. Frydman. {\em Local and bulk 13C hyperpolarization
in nitrogen-vacancy-centred diamonds at variable fields and orientations.} Nature
Comm. {\bf 6}, (2015).
	
\bibitem{king2015room}
J.P. King, K. Jeong, C.C. Vassiliou, C.S. Shin, R.H. Page, C.E. Avalos, H.-J. Wang,
and A. Pines. {\em Room-temperature in situ nuclear spin hyperpolarization from
optically pumped nitrogen vacancy centres in diamond.} Nature Comm. {\bf 6}, (2015).

\bibitem{scheuer2017robust}
J. Scheuer, I. Schwartz, S. M{\"u}ller, Q. Chen, I. Dhand, M.B. Plenio, B. Naydenov,
and F. Jelezko. {\em Robust techniques for polarization and detection of nuclear spin
ensembles.} E-print arXiv:1706.01315
	
\bibitem{salvatier2016}
	J. Salvatier, T. V. Wiecki, Ch. Fonnesbeck.
	\textit{Probabilistic programming in Python using PyMC3.}
	PeerJ Computer Science 2 (2016): e55.
	
\bibitem{hoffman2014}
	M. Hoffman, A. Gelman.
	\textit{The No-U-turn sampler: adaptively setting path lengths in Hamiltonian Monte Carlo.}
	Journal of Machine Learning Research 15.1 (2014): 1593-1623.
	
\bibitem{gaebel2006}
	T. Gaebel, M. Domhan, I. Popa, C. Wittmann, P. Neumann, F. Jelezko, ..., J. Meijer.
	\textit{Room-temperature coherent coupling of single spins in diamond}.
	Nature Physics, 2(6), 408-413. (2006)
	
\bibitem{feroz2009}
	F. Feroz, M. Hobson, M. Bridges.
	\textit{MultiNest: an efficient and robust Bayesian inference tool for cosmology and particle physics.}
	Monthly Notices of the Royal Astronomical Society 398.4 (2009): 1601-1614.
	
\bibitem{hilbe2017}
	J. Hilbe, R. de Souza, E. Ishida.
	\textit{Bayesian Models for Astrophysical Data: Using R, JAGS, Python, and Stan.}
	Cambridge University Press, 2017.
	
\bibitem{flegal2011}
	J. Flegal, G. Jones
	\textit{Implementing MCMC: estimating with confidence.}
	Handbook of Markov chain Monte Carlo, Boca Raton, Florida: Chapman \& Hall/CRC, 175-197. (2011).
	
\bibitem{Footnote3} Due to large deviations in the SNR caused by the Poissoninan photon
detection, we do not expect an exact linear line, but only a trend.

\bibitem{Footnote4} The detection efficiency can be significantly increased to almost 100\% by using single-shot readout.

\bibitem{Footnote_PYMC} Practically the performance of the sampling highly depends on it's initial starting value \cite{flegal2011}. 
Therefore \mbp{an as good as possible} a posteriori (MAP) estimate is used as a starting point. Recently 
tools for probabilistic programming (PP), automatic differentiation frameworks and advances in MCMC methods 
made automatic Bayesian inference on PGMs easy to formulate and perform. The tool used in this work is called 
PyMC3 \cite{salvatier2016}. The framework automatically derives a likelihood function for the model and repeats 
the sampling and evaluation for a a defined upper bound. This reduces implementation effort and make 
quick model changes possible.

\bibitem{Kehayias2017}
	P. Kehayias, et al.
	\textit{Solution nuclear magnetic resonance spectroscopy on a nanostructured diamond chip.}
	arXiv preprint arXiv:1701.01401 (2017).
	
\bibitem{Hincks2017} I. Hincks, C. Granade, and D.G. Cory, {\em Statistical Inference with Quantum Measurements: Methodologies for Nitrogen Vacancy Centers in Diamond.} E-print arXiv:1705.10897

\bibitem{Peck1995} Peck, Timothy L., Richard L. Magin, and Paul C. Lauterbur. {\em Design and analysis of microcoils for NMR microscopy.} Journal of Magnetic Resonance, Series B 108.2 (1995): 114-124.
	
\bibitem{Manson2006}
	N.B. Manson, J.P. Harrison,and M.J. Sellars.
	\textit{Nitrogen-vacancy center in diamond: Model of the electronic structure and associated dynamics.}
	Physical Review B. {\bf 74}, 104303 (2006).
	
\bibitem{supp}
	See supplementary information.
	
\bibitem{SchwartzST+2017} I. Schwartz, J. Scheuer, B. Tratzmiller, S. Müller, Q. Chen, C. Müller, B. Naydenov, 
F. Jelezko and M.B. Plenio. {\em Pulsed Polarisation for Robust DNP.} In preparation

\end{thebibliography}

\end{document}